\documentclass[12pt]{article}

\def\beq{\begin{eqnarray}}
\def\eeq{\end{eqnarray}}

\def\mpl{M_{\rm Pl}}

\def\lsim{\mathrel{\rlap{\lower3pt\hbox{\hskip0pt$\sim$}}
     \raise1pt\hbox{$<$}}}         
\def\gsim{\mathrel{\rlap{\lower4pt\hbox{\hskip1pt$\sim$}}
     \raise1pt\hbox{$>$}}}         

\begin{document}

\begin{titlepage}

\begin{flushright}
{ NYU-TH-03/12/21}

\end{flushright}
\vskip 0.9cm

\centerline{\Large \bf  Graviton Mass or Cosmological Constant?}

\vskip 0.9cm
\centerline{\large Gregory Gabadadze and Andrei Gruzinov}
\vskip 0.4cm
\centerline{\em Center for Cosmology and Particle Physics }
\centerline{\em Department of Physics, New York University, New York,  
NY, 10003}
\vskip 0.9cm

\begin{abstract}

To describe a massive graviton in  4D Minkowski space-time 
one introduces a quadratic term in the Lagrangian. 
This term, however, can  lead to a readjustment or 
instability of the background instead of describing a massive 
graviton on  flat space. We show that for all local 4D
Lorentz-invariant mass terms Minkowski space is unstable. 
The instability can develop in a time scale that is many orders of magnitude 
shorter than the inverse graviton mass. We start with  the Pauli-Fierz (PF) 
term that is the only local mass term with  no ghosts in the 
linearized approximation. We show that nonlinear completions of the 
PF Lagrangian give rise to instability of Minkowski space.
We continue with  the mass terms that 
are not of a PF type.  Although these models 
are known to  have ghosts in the linearized approximations, 
nonlinear interactions can lead  to
background change due to which the ghosts are eliminated. 
In the latter case, however, the graviton perturbations on 
the new background  are not massive. 
We argue that a consistent theory of a massive 
graviton on flat space can be formulated in 
theories with extra dimensions. They require an infinite number of 
fields or non-local description from a 4D point of view.

\end{abstract}

\vspace{3cm}


\end{titlepage}

{\it 1. Introduction.}~ One expects that a massive state
mediates the Yukawa interaction  at distances larger than
its Compton wavelength\footnote{An  exception is a 
massive photon in the Maxwell-Chern-Simons theory in $(2+1)$ dimensions
where the potential is power-like.}. To describe a massive particle 
one adds a quadratic in fields term to the Lagrangian. 
This guaranties that at least in
classical theories with no gravity  large distance interactions are of 
a Yukawa type. Likewise, to describe a massive graviton in  4D Minkowski space-time 
one would introduce a quadratic mass term. However, 
this term could lead to a change of 
the gravitational background, instead of describing  
a massive graviton on  flat space. 
Below we will discuss local Lorentz invariant quadratic 
``mass terms'' for gravity in 4D.  We will argue  that in all
the cases of physical relevance the ``mass term'' leads 
to instability of Minkowski space.  The instability 
can set in within a time scale that is arbitrarily small 
compared to the inverse graviton mass. 
The resulting theory either has no stable vacuum at all, or the original Minkowski 
space is readjusted to a curved background.  Along the way, we draw attention to   
an interesting phenomenon: a theory that has a ghost in the linearized 
approximation, can become ghost free due to nonlinear interactions
that lead to the readjustment of the gravitational background.

We will argue that a natural way to have a massive graviton 
on flat background is to invoke certain theories with infinite-volume 
extra dimensions. The latter have an infinite number of states in the 
spectrum at arbitrarily low energy scale.  From the point of view of 4D
they are nonlocal field theories.

The paper is organized as follows: in section 2 
we start with the PF massive gravity. We show 
that the PF term, and any of its nonlinear polynomial completion,  
gives rise to instabilities of flat space. We find new 
cosmological solutions in empty space that describe instability of the  
Minkowski background and we discuss the time-scale in which the 
instability can set in.

In section 3 we discuss non-PF quadratic terms. These terms are 
traditionally discarded since they give rise to ghosts 
already in the linearized approximation. We show that 
a reparametrization invariant nonlinear completion 
of at least one of these models gives rise to 
a background change. There are no ghosts on a 
new background, however, a graviton does not mediate 
Yukawa potential at large distances.

In the light of our findings, in section 4  
we comment on the strong coupling problem 
in massive gravity. The issue should be addressed 
on a stable (or long-lived metastable) background if such a 
background exists. A conclusion on whether the 
strong coupling problem is present or not depends 
in general on the properties of the background itself.
For 4D PF gravity, however,  we could not find any 
convincing arguments in favor that the theory possesses 
a stable (or very long-lived) ground state in which 
the problem could be studied.

Finally in Section 5 we discuss how a model  
of massive gravity on a flat background 
can be obtained in theories with infinite volume 
extra dimensions. We emphasize certain distinctive features 
that enable these models to accommodate a flat space massive graviton.
A brief summary of main results is given in section 6. 


{\it 2. Pauli-Fierz gravity.}~ In the linearized 
approximation the PF term is introduced  as follows \cite {PF}:
\beq
S^{L}_{PF}\,=\,{\mpl^2\over 8} \,\int & d^4x &
\left ( \partial_\alpha h_{\mu\nu} \partial^\alpha h^{\mu\nu}
-2 \partial_\alpha h_\mu^{\alpha} \partial_\beta h^{\mu \beta}
+2 \partial_\alpha h_\mu^{\alpha} \partial^\mu  h^{\beta}_{\beta}
- \partial_\alpha  h_\mu^{\mu} \partial^\alpha  h^{\nu}_{\nu} \right )\,-
 \nonumber  \\
\,{\mpl^2\,m_g^2 \over 8}\,
\int\,& d^4x & \,\left (h_{\mu\nu}^2\,-\,(h^\mu_\mu)^2\right )\,,
\label{PF}
\eeq
where $m_g$ stands for the graviton mass and $h_{\mu\nu}$
denotes graviton perturbation on a flat background. 
The first term in the parenthesis on the r.h.s. 
of (\ref {PF}) is the linearized Einstein-Hilbert term.

The action (\ref {PF}) describes a consistent theoretical model 
of a free massive spin-2 state with five physical 
degrees of freedom.  This can easily be seen by making 
the reparametrization invariance of this action 
manifest using St\"ukelberg's method. This action could be useful for, e.g., 
a spin-2 glueball  in QCD with $m_g\sim 2$ GeV and 
$\mpl\to M_{\rm QCD}\sim 1$ GeV, however, the action (\ref {PF}) cannot  
describe observable gravity.  This is primarily because of 
the van Dam-Veltman-Zakharov (vDVZ) discontinuity 
\cite {Veltman,Zakharov} (see also \cite {Iwa}), and because (\ref {PF}) 
does not contain nonlinear gravitational interactions that are 
being measured in gravity observables. The non-linearities could 
cure the vDVZ discontinuity problem as well 
\cite {Arkady}.  Therefore, a nonlinear completion of the action 
(\ref {PF}) is needed. However, this in general leads to 
problems \cite {Deser}. The simplest strategy is to 
complete the kinetic term in (\ref {PF}) to a nonlinear 
Einstein-Hilbert term:
\beq
S_m\,=\,-{\mpl^2\over 2} \,\int\,d^4x\,\sqrt{g}\,R(g)\,- 
\,{\mpl^2\,m_g^2 \over 8}\,
\int\,d^4x\,\left (h_{\mu\nu}^2\,-\,(h^\mu_\mu)^2\right )\,,
\label{PFnl}
\eeq
where we define $h_{\mu\nu}\equiv (g_{\mu\nu} -\eta_{\mu\nu})$.
Note that with this definition of $h$ the mass term in the action
(\ref {PFnl}) is regarded as an exact term and not as a 
leading term in a small $h$ expansion. 
Furthermore, higher powers in  $h$  could be  arbitrarily
added to the mass term  since there is
no principle, such as reparametrization
invariance, that could fix the arbitrarity in choosing 
those terms. For definiteness, we can assume that
the  indices in the mass term are raised and lowered by
$\eta_{\mu\nu}$; using $g_{\mu\nu}$ instead,
would result in differences that  appear only in the cubic and higher
orders in $h$ which are ambiguous anyway.

One may attempt to find a more satisfactory than (\ref {PFnl}) 
completion of the PF term by expressing $h_{\mu\nu}$ in terms of the 
invariant curvatures in a certain nonlocal way. 
However, because of the specifics of the PF term this is not 
conceivable.
Indeed, consider the equation of motion that follows from 
variation of (\ref {PF}). Let us take a derivative of 
both sides of the equation.
Since the Einstein tensor is identically conserved, this 
gives a new constraint arising from the mass term.
This is an analog of the Proca condition for massive gauge fields. 
In the case of the PF term the Proca condition reads:
\beq
\partial^\mu h_{\mu\nu} \,=\, \,\partial_\nu h^\alpha_\alpha\,.
\label{proka}
\eeq
An important fact is that for any field that satisfies (\ref {proka})
the Ricci scalar is zero in the linearized approximation. Hence,
the filed $h$ cannot be expressed via the Ricci scalar. Let us now look at 
the Ricci tensor. For the fields that satisfy (\ref {proka}) we find
\beq
R_{\mu\nu}\,=\,{\hat P}_{\mu\alpha}h^{\alpha}_\nu\,,~~~
{\hat P}_{\mu\alpha}\,\equiv\,\partial^2\eta _{\mu\alpha}\,
-\,\partial_\mu\partial_\alpha \,.
\label{proj}
\eeq
Hence, the Ricci tensor and $h$ are related by a 
projector operator ${\hat P}$ which  is not invertible
for general configurations. Therefore, $h_{\mu\nu}$ 
cannot be expressed via the Ricci tensor either. 

We continue with the action (\ref {PFnl}).
Once this completion is adopted problems emerge. 
On a flat background the nonlinear theory (\ref {PFnl}) 
describe a massive spin-2 state with five degrees of freedom 
plus a ghost-like spin-0 state that appears only on a nonlinear level 
\cite {Deser}. The Hamiltonian 
for $h$ is not positive-semidefinite \cite {Deser}. 
This indicates that Minkowski space should be unstable.

At a first glance one might think that the typical time-scale 
for the instability should be of the order of the inverse graviton mass, 
since this is the only new dimensionful parameter in the Lagrangian. 
If this were true, then the theory with the graviton mass
as small as $m_g\sim H_0\sim 10^{-42}$ GeV would have been
almost stable for all the practical purposes. However, as 
we will show shortly, this is not so. Below we derive  exact 
{\it empty-space solutions} of PF gravity  that take the background 
away from Minkowski space and show that the time scale for setting in
this instability can be arbitrarily short.

To exhibit the instability of Minkowski space  
it is enough to focus on the following restricted class 
of metrics:
\begin{equation}
ds^2\,=\,N^2dt^2\,-\,a^2d{\bf x}^2\,,
\label{metric}
\end{equation}
where $N=N(t)$ and $a=a(t)$ are some functions of the time coordinate. 
The Lagrangian for these configurations takes the form (below we 
set $\mpl=1$):
\begin{equation}
L=-N^{-1}a\dot{a} ^2\,-\,m_g^2F\,,
\end{equation}
where $F=F(a,N)$ denotes a general mass term.
From the above Lagrangian we calculate the Hamiltonian 
and find the conserved energy:
\begin{equation}
E\,=\,-N^{-1}a\dot{a} ^2\,+\,m_g^2F\,.
\end{equation}
The function $N$ should also satisfy a constraint 
\begin{equation}
N^{-2}a\dot{a} ^2\,=\,m_g^2\partial _NF\,.
\end{equation}

It is convenient to make a change of variables 
$Ndt\rightarrow dt$ (note that this is not a coordinate transformation 
under which the massive theory is invariant, this  is 
just a formal change of variables used for technical simplifications). 
In terms of the new time variable the constant energy reads 
\begin{equation}
E=-Na\dot{a} ^2\,+\,m_g^2F,
\end{equation}
and the constraint takes the form
\begin{equation}
a\dot{a} ^2\,=\,m_g^2\partial _NF.
\label{constraint1}
\end{equation}

Let us now turn to the PF mass term. For the metric (\ref {metric})
the PF term takes the form: 
$F=(N^2+a^2-2)(1-a^2)/4$. 
The corresponding conserved energy is
\begin{equation}
m_g^2E=-{a^2\dot{a} ^4\over 1-a^2}-{m_g^4\over4}(2-a^2)(1-a^2)\,.
\end{equation}
The latter expression can be rewritten as follows:
\begin{equation}
a^2\dot{a} ^4\,+\,m_g^2(1-a^2)(E+m_g^2(2-a^2)(1-a^2)/4)\,=\,0\,.
\label{aE}
\end{equation}
Minkowski space, that is $a=1$, $N=1$, is certainly 
a solution of the above equation. However, perturbations take 
the solution far away  from Minkowski background, as we 
will see below. 

To see the instability of Minkowski space manifestly 
we need  to study  perturbations for 
both $a$ and $N$ near the point  $a=1$, $N=1$. 
For this,  let us take $E=-m_g^2 \epsilon /2$,
with small positive $0< \epsilon \ll 1$. 
For small $\delta \equiv 1-a$ we find
\begin{equation}
\dot{\delta } ^4\,=\,m_g^4\delta (\epsilon -\delta)\,.
\end{equation}
The above equation describes oscillations of the delta between 
$0$ and $\epsilon$. On the other hand, one can show that
for $\delta\neq 0$  
\begin{equation}
N^2=(\epsilon -\delta )/\delta .
\end{equation}
When $\delta =\epsilon /2$, we have $N=1$. That is the 
solution passes close to the Minkowski region. However, 
at the turning points, $N\rightarrow 0$ or $N\rightarrow \infty$, 
the system moves away from  Minkowski space
\footnote{We would like to point out again that the nonlinear theory lacks 
reparametrization invariance, and, hence, different choices of coordinates 
could lead to different physical spaces. Our interval is defined 
by (\ref {metric}), and what we call Minkowski space corresponds to 
the point $a=1$, $N=1$. Note that geodesic equations 
for matter fields are not modified as compared to the standard GR 
and, therefore, external sources moving along the geodesics
would not distinguish between the coordinate systems. However, 
from the point of view of pure PF gravity, different coordinates can be
physically different.}. Consider the regime when $\delta \ll \epsilon$.
The time dependence of the small $\delta $ takes the form:
$\delta \sim (m_gt)^{4/3}\epsilon ^{1/3}$, and 
$N$ scales as follows: $N^2\sim \epsilon^{2/3}/(m_gt)^{4/3}\gg 1$.
Thus, a small departure  from $a=1$  leads to a 
large deviation from the $N=1$ point (i.e., from Minkowski space).
Note that a typical time scale for the system to complete one cycle
between the turning points is $T\sim \sqrt {\epsilon}/m_g$. The latter 
can be arbitrarily small. Therefore, the instability of Minkowski space 
could develop almost instantaneously. The appearance of a new short time 
scale  in due to the integration constant (i.e., the energy $E$)
which does not enter as a parameter in the Lagrangian.

Actually, one can show that perturbations around Minkowski 
space-time with negative energies exist for arbitrary 
(non-linearly completed) polynomial
mass terms. For non-linear completions of the special form, when the 
mass term in the action is a function 
$f\left (h_{\mu\nu}^2\,-\,(h^\mu_\mu)^2\right )$ 
this was shown in \cite {Deser}. 
This can be generalized for an arbitrary mass term, as 
we have shown it in the Appendix. 

For the PF mass term  there also exists a curious 
``cosmological'' solution. Consider  universe with $E=-m_g^2/2$. 
For this case Eq. (\ref {aE}) simplifies and we get 
\begin{equation}
\dot{a} ^4\,=\,m_g^4(1-a^2)(3-a^2)/4\,.
\label{curious}
\end{equation}
This describes an expanding and then recollapsing universe. 
The early time expansion law $a=m_gt/\sqrt{2}$ corresponds to the 
equation of state $p=-\rho /3$. Note also that the Minkowski space,
$a=1$, $N=1$, is formally a solution of the system (\ref {constraint1}, 
\ref {curious}), however, for small $\delta =1-a \ll 1$
the perturbation of $N$ is huge, $N\sim 1/\sqrt{\delta}$, and the 
corresponding energy is negative. Therefore, small perturbations in $a$ 
move the system from Minkowski space away to a collapsing universe.

\vspace{0.1in}

{\it 3.Non-PF terms.}
We showed above that  Minkowski space is unstable 
for the PF theory. Therefore, there is no reason to prefer the PF term over  
any other non-PF quadratic terms, for which it is known that ghosts 
appear already in the linearized approximation \cite {Neu}. On the other hand, 
choosing non-PF terms one might hope to find a nonlinear 
completion  for  which the ghost will be  eliminated by nonlinear 
interactions. We will discuss this possibility below. 

Let us first start with a general non-PF quadratic term
\footnote{This form of the mass term does not include the case 
when $h_{\mu\nu}^2$ term is absent. However, this should be 
similar to the other generic $a\neq 1$ cases.}
\beq
\,{\mpl^2\,m_g^2 \over 8}\,
\int\,d^4x\,\left (h_{\mu\nu}^2\,-\,a\, (h^\mu_\mu)^2\right )\,,
\label{nonPF}
\eeq
where $a\neq 1$.  In this case the Proca condition
takes the form
\beq
\partial^\mu h_{\mu\nu} \,=\, a \,\partial_\nu h^\alpha_\alpha\,.
\label{prokaa}
\eeq
As a result, the 4D curvature in the linearized theory 
is not identically zero,  $R\sim (a-1) \partial^2 h^\alpha_\alpha$ 
(unlike the case of the PF term).
However, for $a\neq 1$ the term (\ref {nonPF})
gives rise to a ghost. The easiest way to see this is to 
focus  on the scalar  $\phi$ where 
$h_{\mu\nu} =\partial_\mu  \partial_\nu\phi$.
For this scalar the integrand in (\ref {nonPF}) reads:
\beq
(1-a)\left (  \partial^2 \phi \right )^2\,.
\label{ghost}
\eeq
The energy density  that 
follows from (\ref {ghost})
\beq
{\cal E}\,\propto \,(1-a)\left [(\partial_0^2\phi)^2 \,-\, 
(\partial_i^2\phi)^2\right ]\,,
\label{eps}
\eeq
is not positive  definite irrespective of the 
sign of $(1-a)$. In terms of a propagator for $\phi$, 
one finds  a pole with a negative residue -- a ghost.
Is it possible to overcome this inconsistency of the theory?
This question can be given a positive answer 
at least for a certain choice $a=1/2$. This is 
due to a  mechanism  
that we will describe briefly below (similar mechanism was used 
in a higher-dimensional context to stabilize ghosts in Ref. 
\cite {ggg}).  To focus on the main idea   
in as simple terms as possible consider a scalar field 
theory in the absence of gravity:
\beq
{\cal L}\,=\,{\cal G}(\Phi,\chi)\,
\partial^\alpha \Phi \,\partial_\alpha \Phi \,-V(\Phi,\chi),
\label{scalar}
\eeq
where ${\cal G}$ encodes nonlinear interactions of $\Phi$,
its derivatives and/or other fields collectively denoted by $\chi$:
\beq
{\cal G}(\Phi, \chi)\, \equiv \,-\,{1\over 2}\,+\,
{\cal O }\left (\Phi;\, \partial \Phi; \,\chi;\, \partial \chi \right )\,.
\label{phi}
\eeq
The sign of the first term on the r.h.s. of (\ref {phi})
is such that  small perturbations
of  $\Phi$ around $\Phi=0$ are unstable, i.e.,
these perturbations have negative signature kinetic term and 
are ghost-like. However, due to nonlinear interactions
one can change the signature of the kinetic term  
(\ref {phi}). This can be done in a few ways: 

(i) Consider an example 
\beq
{\cal G}(\Phi)\, \equiv \,-\,{1\over 2}\,+\,
{\Phi^2\over v^2}\,.
\label{phi1}
\eeq
Furthermore, let the potential $V$ in (\ref {phi}) take the form:
\beq
V(\Phi)= \lambda (\Phi^2-v^2)^2\,.
\label{scalar1}
\eeq
Then the vacuum solution is $\Phi=v$ and a 
small perturbation $\sigma $ around the  vacuum, 
$\Phi= v+\sigma$, acquires a kinetic term with a positive 
signature\footnote{Note that the phases of the above model 
with $\Phi=0$ and $\Phi=v$ can in general be disconnected 
from each other (superselection sectors), however, this is 
not a matter of our discussions.} 
(as long as $|\sigma| \ll v$):
\beq
\left ( {1\over 2}\,+\,
{2 \sigma \over v}\,+\,{\sigma^2 \over v^2} \right)\,
(\partial\sigma)^2+...\,.
\label{pos}
\eeq

(ii) The second example is similar to the first one, 
but it is due to higher derivatives. Consider 
\beq
{\cal G}(\chi)\, \equiv \,-\,{1\over 2}\,+ \,{\partial^2 
\chi \over v^3}\,. 
\label{phi2}
\eeq
Suppose that for certain dynamical reasons  
the $\chi$ field develops the following condensate:
\beq
\langle \partial^2 \chi \rangle \,= \,v^3\,.
\label{HD}
\eeq
This condensate leads to the ``signature change''
for the kinetic term of the $\Phi$ 
field and small perturbations of the $\Phi$ 
field about the correct vacuum state will have a positive 
sign of energy. 

(iii) Finally, nonlinear interactions of a single tensor 
field could be a reason for the elimination of the ghost.
Below we will discuss such a mechanism for a graviton.
For this we will restrict ourselves  to the  case $a=1/2$ 
in (\ref {nonPF}).  This choice is somewhat
special for reasons that will become clear shortly.
We will also comment on the other $a\neq 1$ cases below. 

Thus, we consider the linearized action:
\beq
S^{L}_{gPF}\,=\, {\mpl^2\over 8} \,\int & d^4x &
\left ( \partial_\alpha h_{\mu\nu} \partial^\alpha h^{\mu\nu}
-2 \partial_\alpha h_\mu^{\alpha} \partial_\beta h^{\mu \beta}
+2 \partial_\alpha h_\mu^{\alpha} \partial^\mu  h^{\beta}_{\beta}
- \partial_\alpha  h_\mu^{\mu} \partial^\alpha  h^{\nu}_{\nu} \right )\, -
\nonumber  \\ \,{\mpl^2\,m_g^2 \over 8}\,
\int & d^4x & \left (h_{\mu\nu}^2\,-\,{1\over 2}\,
(h^\mu_\mu)^2\right )\,.
\label{gPF}
\eeq
In the quadratic approximation the above action can be 
rewritten as:
\beq
S^{L}_{gPF}\,=\,{\mpl^2\over 8} \,\int & d^4x &
\left ( \partial_\alpha {\tilde h}_{\mu\nu} \partial^\alpha 
{\tilde h}^{\mu\nu}
-2 \partial_\alpha {\tilde h}_\mu^{\alpha} \partial_\beta 
{\tilde h}^{\mu \beta}
+2 \partial_\alpha {\tilde h}_\mu^{\alpha} \partial^\mu 
{\tilde h}^{\beta}_{\beta}
- \partial_\alpha  {\tilde h}_\mu^{\mu} \partial^\alpha  
{\tilde h}^{\nu}_{\nu} \right )\,- 
\nonumber  \\ \,{\mpl^2\,m_g^2 \over 2}\,
\int & d^4x & \left (  \,-1 -{{\tilde h}\over 2} + {1\over 4} 
\left ({\tilde h}_{\mu\nu}^2\,-\,{1\over 2}\,
({\tilde h}^\mu_\mu)^2\right )\right )\,,
\label{gPF1}
\eeq
where ${\tilde h}_{\mu\nu}\equiv h_{\mu\nu}-\eta_{\mu\nu}$, 
and we used the relation:
\beq
{1\over 4} \left (h_{\mu\nu}^2\,-\,{1\over 2}\, (h^\mu_\mu)^2\right )\,
=\,-1 -{{\tilde h}\over 2} + {1\over 4} 
\left ({\tilde h}_{\mu\nu}^2\,-\,{1\over 2}\,
({\tilde h}^\mu_\mu)^2\right )\,.
\label{dstran}
\eeq
We can imagine that the action (\ref {gPF1}) is 
our starting point in which matter couples to ${\tilde h}$
in a conventional way. The actions (\ref {gPF}) and (\ref{gPF1}) 
describe a free massive spin-2 state plus 
a massive spin-0 ghost. This could be seen by calculating a 
one-particle exchange amplitude between two conserved sources
$ T_{\mu\nu}$ and $T^{\prime\,\mu\nu}$. The momentum space 
amplitude of the linearized theory contains the following terms 
\beq
{ T_{\mu\nu}  T^{\prime\,\mu\nu}\,-\,{1\over 3}\, {T}
\, {T}^{\prime} \over \,m_g^2 -\,p^2 -\,i\epsilon }\,-
{1\over 6}\, { {T} \, {T}^{\prime} \over m_g^2 -\,p^2 -\,i\epsilon }\,,
\label{ag}
\eeq
with $p^2$ being the transfer-momentum square. 
The first term corresponds to an exchange of a massive spin-2 state
while the second term gives rise to a repulsive interaction
due to a massive spin-0 ghost.
Therefore, the model (\ref {gPF}), (or (\ref {gPF1})) 
as it stands, cannot be a consistent theory of gravity.

Being motivated by the scalar field example discussed above 
we will  add new  terms  to (\ref {gPF}) and (\ref {gPF1}) 
to eliminate the ghost. This procedure, as we will see, 
also eliminates the longitudinal polarizations of a massive graviton,
and leads to a theory of a massless graviton on a curved 
background. Thus, we expect that 
\beq
\left (h_{\mu\nu}^2\,-\,{1\over 2}\,
(h^\mu_\mu)^2\right )\,+\,V(h; \eta_{\mu\nu})\,,
\label{add}
\eeq
can describe a theory with no ghosts for certain choices of $V$.
The key observation is that 
\beq
\sqrt {|{\rm det} h_{\mu\nu}|}\,=\,\left (h_{\mu\nu}^2\,-\,{1\over 2}\,
(h^\mu_\mu)^2\right )\,+\, V(h_{\mu\nu}-\eta_{\mu\nu})\,,
\label{expansion}
\eeq
where $V$ is a known polynomial of its argument.
The above relation can be established by using an identity 
$$ \sqrt {|{\rm det}h_{\mu\nu}|}\equiv \sqrt {|{\rm det}(\eta_{\mu\nu}
+h_{\mu\nu}-\eta_{\mu\nu})|},  $$ 
and formally expanding it 
in powers of $h_{\mu\nu}-\eta_{\mu\nu}$:
$$   
\sqrt {|{\rm det}(\eta_{\mu\nu}
+h_{\mu\nu}-\eta_{\mu\nu})|}\,=\,{\rm exp}\left ({1\over 2}{\rm Tr}
\sum_{n=1}^{\infty} {(-1)^{n+1} \over n}(h_{\mu\nu}-\eta_{\mu\nu})^n
\right )\,. 
$$
It is certainly true that $V$ in (\ref {expansion}) contains an 
infinite number of constant, linear, quadratic and higher powers 
of $h$, nevertheless, the above procedure is a nonlinear 
completion for the action (\ref {gPF1}) that is written in terms of 
the variable  ${\tilde h}_{\mu\nu}$, since
the function $V$ contains only cubic and higher powers 
in ${\tilde h}_{\mu\nu}$.
The polynomial terms  in $V({\tilde h})$ trigger 
the background change for the ghost field 
by a mechanism similar to the one described above. The fact that 
this is the case is easy to understand without 
term-by-term calculation of $V$. This is because  the resulting 
nonlinear theory can be written in a simple way
\beq
S\,=\,- {\mpl^2\over 2} \,\int\,d^4x\,\sqrt{h}\,
\left ( R(h)\,+\,2\, m_g^2 \right )\,. 
\label{ds}
\eeq
The above functional  is nothing but the action of a theory 
with nonzero cosmological 
constant equal to $m_g^2$. It admits solutions with curved background but
does not admit flat solutions. The spectrum of the theory on the 
curved background (either de Sitter  or anti-de Sitter) has no ghosts.
The graviton in (\ref {ds}), unlike a 4D massive spin-2 state, propagates 
two physical degrees of freedom.

The results obtained above could be also  understood in the following way: 
Let us start with the Einstein-Hilbert action with a nonzero cosmological
constant (\ref {ds}).  Let us expand this action  {\it formally} 
around a flat background, $\eta_{\mu\nu} +{\tilde h}_{\mu\nu}$.
Note that we are expanding around a background that is {\it not} 
a solution of the equations of motion. Because of this we should 
anticipate  certain 
inconsistencies to emerge. As we will see, the way the inconsistencies 
appear is very instructive, so we continue with our expansion.
We truncate this expansion  at the quadratic
order in ${\tilde h}$ (the covariant derivative in this expansion is
just a simple derivative).  The resulting theory is (\ref {gPF1}),
that has the quadratic ``mass term'' in ${\tilde h}$, 
the linear term, and the constant term.
We regard the resulting model as a certain free theory
of ${\tilde h}$.  Minkowski space, i.e., ${\tilde h}=0$, is 
certainly not a solution of the above linearized theory
(this can also be understood as 
impossibility to obtain a cosmological term starting from flat background
and considering consistent self-coupling requirements for the 
linearized action \cite {Deser1}.). 
However, it is remarkable that in the  linearized theory (\ref {gPF1}), 
the  constant, linear and  quadratic non-derivative terms
can be rearranged as a non-PF  term with $a=1/2$ by a formal change of 
variables (see action (\ref {gPF})). The latter is a non-PF ``mass term'' 
for $h$ on a flat background! It has a ghost in spite of the fact that 
the original nonlinear theory (\ref {ds}) was ghost free.

Two important comments are in order. 

(1) The above derivation applies to the $a=1/2$ case only.
The question is whether the same conclusions remain valid for 
any other $a\neq 1$ case. It is certainly true that 
the linearized theory is unstable (has ghosts) for any $a\neq 1$.
Therefore, to make sense of such models the ``signature change''
have to take place. If so, the background will also be changed.
Then, we come  to a similar conclusions --
either these models are inconsistent, or they describe curved space,
but none of these models can describe a massive graviton on a flat space.

(2) So far we have been  dealing with the 
classical effects only. However, quantum corrections 
can be important in discussing  the issues of massive gravity.
Let us start with the PF mass term again. This term is set
in a classical theory by  adjusting the coefficients
of the $h_{\mu\nu}^2$ and $(h^\alpha_\alpha)^2$ terms
to be equal. However, there is no reason for 
quantum {\it gravitational} loops to preserve this 
condition after the appropriate 
wave-function renormalization 
is performed. These coefficients 
are different in quantum theory and one in general 
is back to the $a\neq 1$ case.

A question arises why the infinite number 
of terms in $V$ that we add in (\ref {add},\ref {expansion})
are stable w.r.t. quantum corrections. The answer 
is that the reparametrization invariance of the 
complete theory (\ref {ds}) protects these terms from being 
renormalized.  Therefore the procedure 
described for the $a=1/2$ case is stable under loop corrections.
These corrections just renormalize the wave-function $h$,
Newton's constant $G_N\equiv 1/8\pi \mpl^2$, 
``graviton mass'' $m_g^2$ (i.e., the cosmological constant),
and give rise to higher derivative terms. 

How do these arguments change for the other $a\neq 1$ models?
It is clear that unless the other $a\neq 1$ models also have a 
reparametrization invariant completion, similar to that of the 
$a=1/2$ case, any finely adjusted nonlinear addition 
to  these models  will  in general be destroyed by 
gravitational loop effects. 

{\it 4. On the strong coupling problem in massive gravity.}~
It has been known for some time that perturbative expansion in 
$G_N$ breaks down  in nonlinear diagrams at a scale 
that is parametrically lower than the UV cutoff of the theory 
\cite{Arkady} (see also \cite {DDGV}). 
This can be understood as a consequence of strongly interacting 
longitudinal modes of massive graviton \cite {AGS}.
At the classical level, the calculations can still be 
performed by means of resummation of the 
tree-level perturbation theory in $G_N$, or by using 
a perturbative expansion in a different parameter 
\cite {Arkady},\cite{DDGV}. However, the question 
whether the same can or cannot be done in full quantum 
PF theory  remains open. If the resummation is not possible in 
the quantum PF theory, then there will appear higher-derivative 
operators in the theory that are suppressed by a phenomenologically 
unacceptable low scale  \cite {AGS}.  

The above results are obtained 
by considering perturbative expansion on a flat background. 
However, as we discussed above, the Minkowski background is 
unstable in PF gravity. Moreover, the instability of Minkowski space 
can set in within a time scale that can be arbitrarily short. 
Therefore, to understand whether the problem is 
truly present in the PF theory, the issue should be studied  
on a stable (or a long-lived metastable) ground state, 
if such a state exists. At the moment the existence 
of such a ground state is not obvious.
If such a state does not exist, then the nonlinear 
version of the PF gravity should be discarded as an inconsistent model. 
The present work has nothing new to add in this regard, all we have shown
is that the Minkowski space is certainly not a candidate 
for such a background. On the other hand, if some stable curved 
ground state exists, then the vDVZ discontinuity and
the strong coupling problems could in principle 
be cured by the background curvature effects 
\cite {Ian,Porrati,AGS}.
 
It is also instructive to mention in this regard different solutions of PF 
gravity that exist in the literature. One starts 
with an empty space and puts a static and 
spherically symmetric source in it. In the linear theory
(\ref {PF}) this source produces a static potential on a flat space 
that has the Yukawa behavior at infinity. However, all this changes in 
the nonlinear  theory (\ref {PFnl}) where we look for 
a spherically symmetric and static solution of  nonlinear 
equations. Moreover, we require that 
the  solution  gives rise to a $1/r$ potential for distances 
$r\ll m_g^{-1}$, and the ${\rm exp} (-m_g r)/r$ potential 
at larger scales. It has been known for some time \cite {cimento,Salam} 
that the solutions of massive gravity in the above two asymptotic 
regimes are hard to match together.
Moreover, recent numerical studies \cite {Kogan} show explicitly 
that the  matching is possible only at the expense of introducing 
a naked singularity at a finite proper distance from a completely  regular 
source. This is certainly unacceptable.  
However, there exist solutions \cite {Salam} 
for which the potential is similar to that of a 
de Sitter-Schwarzschild metric in the static coordinate system
(in that system $g_{00} = 1 - r_g/r -\Lambda r^2$, and $g_{rr}=1/g_{00}$)
\footnote{More precisely, the solution in Ref. \cite {Salam} 
was  found in a different coordinate system in which the off-diagonal 
terms in the metric are not zero. The above solution is reducible
to the static-patch dS-Schwarzschild solution by a formal 
change of coordinates. However, since the reparametrization invariance 
is absent, these two coordinate systems are not physically equivalent.
Nevertheless, we will refer to these metrics as 
dS-Schwarzschild solutions keeping in mind the  above 
disclaimer.}. These solutions also exist 
in the absence of the source, i.e., when $r_g=0$. 
The de Sitter curvature $\Lambda$ is 
determined by  the graviton  mass and a certain integration constant
$\Lambda=(m_g u)^2$, where $u>3/4$. The presence of this arbitrary 
integration constant is reminiscent of an 
arbitrary constant $E$ in the solutions found in section 2.  
Furthermore, unlike the solutions found in the previous section, 
these solutions have a smooth limit as $m_g\to 0$ \cite {Salam}.
The  solution can be  interpreted as follows. 
The {\it gravitational mass} term itself 
acts as a source for gravity and produces effects that are 
somewhat similar to those of a cosmological constant.
An open question remains  whether the dS-Schwarzschild solution itself is 
stable w.r.t. small perturbations. If it is stable and its curvature is 
bigger than $m_g^2$, than there is neither the 
vDVZ nor the strong coupling problems in this case \cite {Ian,Porrati,AGS}.
However, irrespective of whether the curved background is stable or not, 
our main conclusion holds unchanged -- the PF mass term at best 
leads to change of the background, but it in no way  describes a 
flat space massive graviton.

{\it 5. How do extra dimensions help?}~~ 
In a conventional compactifications of theories with extra dimensions 
one obtains a massless graviton that is 
interacting with an infinite number of 
massive spin-2 states. In the linearized approximation 
the mass terms for each of these massive spin-2 KK modes 
have the PF form. As we argued in section 2, flat space  
is unstable for any nonlinear 
completion of the PF mass term. On the other hand,
the original higher dimensional theory is a reparametrization 
invariant model and can be shown to have no instabilities 
of the type obtained in section 2.
The resolution of this seeming contradiction is in the fact that 
one gets an infinite number of massive spin-2 KK states upon 
compactification and truncation of this tower to any finite order
leads to inconsistencies \cite {Nappi},\cite {Duff}.
The manifest reparametrization invariance of a higher dimensional theory
is a convenient book-keeping tool to utilize to see these properties.
The  reparametrization invariance at 
each KK level is maintained on the same KK level only in the 
linearized approximation. Nonlinear effects mix different 
KK levels under the coordinate transformations \cite {Nappi},\cite {Duff}. 
Hence, the consistency of the theory is achieved by means of an 
infinite number of four-dimensional reparametrization invariances. 
Any truncation of  the theory to a finite number of massive spin-2 
fields leads to an explicit breakdown of all the massive gauge invariances, 
including the ones that correspond to the massive fields that are 
retained in the low energy description. As a result, in the truncated 
theory the problems of PF gravity will arise. Therefore, a 
consistent theory should maintain all the infinite number of fields.

In conventional compactifications one obtains a massless graviton.
In this case, large distance gravity is indistinguishable from 
4D general relativity. Our goal, however, is to present 
a model of a massive graviton (with no massless mode). 

A {\it generally covariant} model that  shares many properties 
of massive gravity, but retains all the attractive features of a 
higher dimensional reparametrization invariant  
theory is the DGP model \cite {DGP}.
In five-dimensional context it described a metastable graviton
with no mass. In higher dimensional generalizations 
of the DGP model \cite {DG,MP} the graviton has 
an effective mass that is much larger that its width \cite{Wagner,DHGS}.
Thus, the model introduces  a reparametrization invariant 
``mass term'' for a graviton. Such models have string theory 
realization \cite {Ignat} (see Refs. \cite{DGS} -- \cite {Gruzinov} 
for interesting cosmological and astrophysical studies).

Gravitational dynamics encoded in the model 
can be inferred  both  from the four-dimensional 
as well as $(4+N)$-dimensional standpoints.
 From the 4D perspective,  gravity on the brane is mediated
by an infinite number of the Kaluza-Klein  modes that
have no mass gap. Under  conventional circumstances
(i.e.,  with no  brane kinetic term) this would
lead to higher-dimensional interactions. However,
the large  Einstein-Hilbert term on the brane 
suppresses the wave functions of  heavier KK modes,
so that in effect they  do not participate  in the
gravitational interactions  on the brane at observable distances
\cite {DGKN}.  Only  light KK modes, with masses $m_{KK}\lsim m_g
\sim 10^{-42}$ GeV, remain essential,
and they collectively act  as an effective 4D graviton with
a typical mass  of the order of $m_g$.
At present, the $N\geq 2$ DGP models \cite {DG,MP} 
seem to be the only consistent model of a massive graviton on flat space
(the $N=1$ model is not massive).
This model has no ghosts in the linearized theory \cite {MP, GS}
and possesses a reparametrization invariant 
nonlinear action. Because of this, unlike nonlinear PF gravity,
the ghosts do not appear in the nonlinear 
theory and instabilities of the PF gravity are not present. 

The above models  evade the problems of the 4D PF theory
because at any low-energy scale they contain an infinite number of KK 
gravitons with no mass gap. In other words, 
these models can be thought of as nonlocal models from the 4D
point of view \cite {DGS}\footnote{  
As was proposed in \cite {ADDG}, non-localities postulated in pure 
4D theory (for whatever reasons they might appear),  
can solve an ``old'' cosmological constant problem \cite {ADDG}, 
and give rise to  new mechanisms  for the present-day acceleration 
of the universe~\cite{ADDG}.}. Indeed a massive graviton in 4D can be 
described by a nonlocal equation   
$(1+{m_g^2/ \nabla^2})G_{\mu\nu}=T_{\mu\nu}+... ,$ 
where dots stand for some other terms that are 
needed to restore the Bianchi identities.

We also note that in (2+1) dimensions a unitary and causal 
theory of a massive graviton is topologically massive gravity \cite {Top}.

{\it 6. Conclusions.} ~~Summarizing, the PF term is the only 
quadratic term that has no ghosts 
in the linearized theory. Any polynomial 
nonlinear completion of this term, however, 
gives rise to instabilities. We found empty space 
solutions that manifestly show the instability of Minkowski space in
PF gravity. Therefore, there is no reason to restrict ourselves 
to the PF term and one might start with a non-PF quadratic terms. 
The latter have ghosts already in the linear theory. Nevertheless, 
the ghosts can be eliminated by  higher derivative terms 
via the background rearrangement. 
The resulting nonlinear theory has no classical instabilities, 
however, it does not describe a graviton mediating Yukawa interactions
on flat 4D space. It is very likely that out of all candidates for 
``massive'' local 4D theories, only one has a nonlinear completion that 
is radiatively stable (the $a=1/2$ case). 
However, in this case the ``mass'' term is 
nothing but the cosmological term. A natural way to account for a 
massive graviton on flat space is to invoke theories with extra dimensions.
The latter evade the problems of the 4D massive gravity because 
they are non-local theories from the 4D point of view. 

\vspace{0.2in}

{\it  Acknowledgments.} ~~ We would like to thank Jose Blanco-Pillado 
and Alberto Iglesias for useful and stimulating discussions, and Stanley 
Deser and Arkady Vainshtein for useful comments on the manuscript. The work of AG 
is supported by the David and Lucile Packard Foundation. 

\vspace{0.4in}

\section{Appendix}

We will consider non-linear oscillations of a  massive 
graviton condensate, meaning that all components of the metric 
are functions of $t$ only. It is convenient to use the ADM 
parametrization of the metric:
\begin{equation}
ds^2=N^2dt^2-\gamma _{i j}(dx^{i }+N^{i}dt)
(dx^{j }+N^{j }dt)\,.
\end{equation}
Then the Lagrangian reads
\begin{equation}
L={1\over 2}\sqrt{\gamma } N^{-1}\left( \dot{\gamma }_{ij}
\dot{\gamma }^{ij}-(\dot{\gamma }_{j }^{j })^2 \right)
-m_g^2F,
\end{equation}
where $F(N,N_{i},\gamma _{ij})$ represents 
a general  non-linearly completed Pauli-Fierz term. The $N_{j}$ 
constraints simply remove the $N_{j}$ dependence of $F$:
\begin{equation}
F\rightarrow F(N,\gamma _{ij}).
\end{equation}

It remains to remove the $N$ constraint. Diagonalizing the 3-metric, 
$\gamma _{ij} = {\rm diag}(e^{2a},e^{2b},e^{2c})$, and writing 
$N^2=e^{2d}$, we obtain a simple Lagrangian
\begin{equation}
L=-e^{a+b+c}e^{-d}(\dot{a}\dot{b}+\dot{a}\dot{c}+\dot{b}\dot{c})
-m_g^2F(a,b,c,d).\end{equation}
The quadratic part of $F$ is the Pauli-Fierz quadratic form 
\begin{equation}
F=-(ab+ac+bc)-d(a+b+c)+G,
\end{equation}
and $G$ contains cubic and higher order terms. This is easy to check. 
In the linearized approximation, the Lagrangian becomes
\begin{equation}
L=-(\dot{a}\dot{b}+\dot{a}\dot{c}+\dot{b}\dot{c})+m_g^2((ab+ac+bc)+d(a+b+c)).
\end{equation}
The $d$-constraint gives $a+b+c=0$, leading to
\begin{equation}
L={1\over 2}(\dot{a}^2+\dot{b}^2+\dot{c}^2)-{1\over 2}m_g^2(a^2+b^2+c^2)\,,
\end{equation}
which describes constrained positive-energy harmonic oscillators of mass 
$m_g$.

In the generic nonlinear case, the $d$-constraint gives
\begin{equation}\label{nlconstr}
e^{a+b+c}e^{-d}T-m_g^2\partial _dF=0,
\end{equation}
where
\begin{equation}
T\equiv \dot{a}\dot{b}+\dot{a}\dot{c}+\dot{b}\dot{c}.
\end{equation}
The energy is
\begin{equation}\label{energy}
E=-e^{a+b+c}e^{-d}T+m_g^2F=m_g^2(F-\partial _dF),
\end{equation}
where $d=d(a,b,c,T)$ from (\ref{nlconstr}).

To show that the energy can be negative for arbitrary $F$, 
we assume that  $a,b,c,d,T$ are infinitesimals of the same order. Then we 
can linearize both the constraint equation (\ref{nlconstr}) and 
the energy expression (\ref{energy}). The constraint equation is
\begin{equation}\label{constr0}
T=-m_g^2(a+b+c)\,.
\end{equation}
Our assumption that $a,b,c,d,T$ are 
infinitesimals of the same order will be correct only if we 
choose $a,b,c,T$ that satisfy (\ref{constr0}). Then the energy is 
\begin{equation}
E=-T,
\end{equation}
which is not positive semi-definite.

\end{document}